\documentclass[prl,twocolumn,floatfix,showpacs,preprintnumbers, 
               superscriptaddress,amsmath,amssymb]{revtex4}
\usepackage{graphicx}
\usepackage{epsf}  
\usepackage{ucs}
\usepackage[utf8]{inputenc}

\def\k{{\bf k}}

\begin{document}
\title{Strongly enhanced shot noise in 
%symmetrically coupled 
chains of quantum dots}

\author {Jasmin Aghassi} 
\affiliation{Forschungszentrum Karlsruhe, Institut f\"ur Nanotechnologie,
76021 Karlsruhe, Germany}
\affiliation{Institut f\"ur Theoretische Festk\"orperphysik,
Universit\"at Karlsruhe, 76128 Karlsruhe, Germany}

\author {Axel Thielmann} 
\affiliation{Forschungszentrum Karlsruhe, Institut f\"ur Nanotechnologie,
76021 Karlsruhe, Germany}
\affiliation{Institut f\"ur Theoretische Festk\"orperphysik,
Universit\"at Karlsruhe, 76128 Karlsruhe, Germany}

\author {Matthias H. Hettler} 
\affiliation{Forschungszentrum Karlsruhe, Institut f\"ur Nanotechnologie,
76021 Karlsruhe, Germany}

%\author {J\"urgen K\"onig} 
%\affiliation{Institut f\"ur Theoretische Physik III, Ruhr-Universit\"at
%             Bochum,  44780 Bochum, Germany}
\author {Gerd Sch\"on} 
\affiliation{Forschungszentrum Karlsruhe, Institut f\"ur Nanotechnologie,
76021 Karlsruhe, Germany}
\affiliation{Institut f\"ur Theoretische Festk\"orperphysik,
Universit\"at Karlsruhe, 76128 Karlsruhe, Germany}

\date{\today}

%%%%%%%%%%%%%%%%%%%%%%%%%%%%%%%%%%%%%%%%%%%%%%%%%%%%%%%%%%%%%%%%%%%%%%%%%%%%%%
\begin{abstract}
We study charge transport through a chain of quantum dots.
The dots are fully coherent among each other and weakly coupled to
metallic electrodes via the dots at the interface, thus modelling a
molecular wire. 
%The current-voltage characteristics as well as the 
%current noise are evaluated to first-order perturbation  
%theory in the coupling to the electrodes.
% For a fully symmetric Hamiltonian of three quantum dots 
If the non-local Coulomb interactions dominate over the inter-dot
hopping we find  strongly enhanced shot noise 
%over a wide range of parameters,
 above the sequential tunneling threshold. 
The current is not enhanced in the region of
enhanced noise, thus rendering the noise  super-Poissonian.
In contrast to earlier work this is achieved even in a 
fully symmetric system. 
The origin of this novel behavior lies in a 
competition of "slow" and "fast" transport channels that are formed due 
to the differing non-local wave functions and total spin of the states
participating in transport.
This strong enhancement may allow direct experimental
detection of shot noise  in a chain of lateral quantum dots.
\end{abstract}

\maketitle
%%%%%%%%%%%%%%%%%%%%%%%%%%%%%%%%%%%%%%%%%%%%%%%%%%%%%%%%%%%%%%%%%%%%%%%%%%%%%%

{\bf Introduction}-- 
Shot noise in mesoscopic devices such as point contacts and quantum dots
has been the subject of intensive research  within the last decade
\cite{blanter}. Recently the interest has been intensified 
both theoretically 
\cite{loss,ferro_leads,bulka,elattari,thielmann,haug-kiesslich,belzig1,thielmann_co} 
and experimentally \cite{safonov,nauen,fujisawa} 
when the importance of shot noise for quantum computing,
nanomechanical systems and molecular devices had been
recognized. However, experiments on these systems are difficult to
perform, since 
one needs to detect the shot noise over the background of $1/f$- noise 
caused by fluctuations in the physical environment and 
measurement equipment. So far even for lateral quantum dots in semiconductor
heterostructures, with far reaching control of system parameters and
advanced low temperature setups, no direct shot noise measurements have 
been reported \cite{fujisawa}.

In an array of quantum dots with Coulomb interactions, 
both the current $I$ and the shot noise $S$ depend on details of the 
electronic spectrum of the  many-body states of the coupled system, as well as 
the coupling strengths of these states to the 
electrodes \cite{bulka,thielmann}.
Thus the combined measurements of current and shot noise provide a
'spectroscopic' tool to gain information about the level structure.
For 'local' systems such as 
single dots or two parallel dots (which can be considered as a
single dot with two or more single-particle levels) the shot noise 
is known to be mostly sub-Poissonian ($S < 2 e I$. 
Super-Poissonian noise is predicted
in asymmetric situations, with level- or spin-dependent couplings to the
quantum dot \cite{ferro_leads,bulka,thielmann},
or sometimes in the Coulomb blockade region \cite{belzig1,belzig2}.
Similar considerations  hold for two dots in series \cite{jasmin-2qd}.

In this letter we show that starting with a chain of three dots
the noise may become super-Poissonian 
even if the dot-electrode couplings are fully symmetric 
(spatial and spin symmetry). 
The  shot noise is  greatly enhanced  {\it in  absolute  magnitude} due to a
strong competition between "slow" and "fast" transport channels, 
if the non-local Coulomb repulsion dominates the inter-dot hopping.
As there is no simultaneous enhancement of the tunneling current, the shot
noise becomes super-Poissonian. 
Such behavior can be achieved involving only electronic degrees of freedom,
and over a large bias region above the sequential tunneling threshold.
%in contrast to the work in Refs. \onlinecite{belzig1,belzig2}.
The strong enhancement of the shot noise in a chain of lateral quantum dots
should allow its direct experimental detection. 

{\bf Model and technique}--
We consider a series of quantum dots, each  with one spin-degenerate
level. Including hopping between the dots as well as intra-dot and
inter-dot (nearest neighbor) Coulomb interactions we arrive at the    
Hamiltonian
$\hat H = \hat H_{\rm L} + \hat H_{\rm R} + \hat H_{\rm dots} + 
 \hat H_{\rm T,L} + \hat H_{\rm T,R}$
with
\begin{eqnarray}
&& \hspace*{-0.8cm} \hat H_{r} = \sum_{\k \sigma}\epsilon_{\k} a_{\k
  \sigma r}^{\dag} a_{\k \sigma r},\,\,
\hat H_{{\rm T},r}= \sum_{i \k \sigma}(t^r_i a_{\k \sigma
            r}^{\dag} c_{i \sigma} 
            + h.c.), \nonumber \\
&& \hspace*{-0.8cm}\hat H_{\rm dots} =\sum_{i \sigma} \epsilon_{i} n_{i \sigma}  
  - t\sum_{<ij> \sigma} 
( c_{i \sigma}^{\dag}c_{j \sigma} +h.c. ) \nonumber \\  
&& \hspace*{+0.2cm} + \,U\sum_i n_{i\uparrow}n_{i\downarrow}  +U_{nn}\sum_{<ij> \sigma \sigma'} n_{i\sigma}n_{j\sigma'} ,
%            \frac{\Delta_{ex}}{2}\sum_{\sigma,\sigma',l\neq l'}  
%            c_{l \sigma}^{\dag} 
%            c_{l' \sigma'}^{\dag}c_{l \sigma'}c_{l' \sigma}, \\
%&&\hat H_{{\rm T},r}= \sum_{i k \sigma}(t^r a_{k \sigma
%            r}^{\dag} c_{i \sigma} 
%            + h.c.),
\label{hamilton}
\end{eqnarray}
where $i=1 \ldots N$ and $r={\rm L},{\rm R}$. 
Here, $\hat H_{\rm L}$ and $\hat H_{\rm R}$ model the non-interacting
electrons with density of states $\rho_e$ (assumed as a constant)
%$= \sum_{\k} \delta(\omega -\epsilon_{\k})$ 
in the left and right electrode described by the 
Fermi operators $a_{\k \sigma r}^{\dag}, a_{\k \sigma r}$.
The "artificial molecule" term
$\hat H_{\rm dots}$ describes the series of dots
with Fermi operators $c_{i\sigma}^{\dag}, c_{i\sigma}$,
on-site energies 
$\epsilon_{i}$ and nearest neighbor hopping $t$.
%,and Coulomb interaction on the molecule
% are Fermi operators for the dot levels, and $n_{i\sigma}=c_{i
% \sigma}^{\dag}c_{i \sigma}$ is the number  operator). 
$U$  and $U_{nn}$  are the strength of 
the intra-dot and  nearest neighbor inter-dot Coulomb repulsion.
%Other terms could be considered by much more
%elaborate models, as done in Ref.~\onlinecite{hettler_prl} 
%for computation of the \iv .
%For the  effects on the shot noise that we wish to study, 
%the above simple molecule model suffices.
Tunneling between the leads and dots is modeled by 
$\hat H_{{\rm T},{\rm L}}$ and $\hat H_{{\rm T},{\rm R}}$.
We restrict ourselves to situations
where the leads couple only to the one adjacent dot.
The coupling strength is characterized by 
%the intrinsic line width 
$\Gamma^r_i = 2 \pi |t^r_i|^2 \rho_e$, where $t^r_i$ are the
tunneling matrix elements.

We are interested in transport through a chain of quantum dots, 
in particular in the
current $I$ and the (zero-frequency) current noise $S$.
They are related to the current operator 
$\hat I = (\hat I_{R} - \hat I_{\rm L})/2$, with $\hat{I}_r = -i(e/\hbar) 
\sum_{i \k \sigma} \left( t^r_i a_{\k \sigma r}^{\dag} c_{i\sigma} - h.c.\right)$ 
being the current operator for electrons tunneling into lead $r$, by 
$I = \langle \hat{I} \rangle$ and 
\begin{equation}
S = \int_{-\infty}^{\infty} dt \langle \delta \hat{I}(t) \delta \hat I(0) 
+ \delta \hat{I}(0) \delta \hat I(t) \rangle
\end{equation}
where 
$\delta \hat I(t)=\hat I(t)-\langle \hat I \rangle$.

%\section{Diagrammatic Technique}
For the calculation of the current $I$ and current noise $S$, we use the
diagrammatic technique on the Keldysh contour developed in 
Ref.~\onlinecite{diagrams}
and expanded for the description of the noise in
Ref.~\onlinecite{thielmann}. In this approach the dot array ($H_{\rm
  dots}$)  is treated exactly  
%quantum mechanical coherent way 
as an interacting many-body system, but is perturbatively (incoherently) 
coupled to the electrodes. We diagonalize the dot Hamiltonian 
$H_{\rm dots}$ and compute the transition rates $W_{\chi\chi'}$ 
(forming a matrix $\bf W$)
between two eigenstates $\chi$ and $\chi'$ in first order perturbation theory
in the coupling strengths $\Gamma^r_i$. 
%The bold face indicates matrix notation related to the eigenstate
%labels $\chi$
%Note that for an $N$ dot system $4^N$ different eigenstates $\chi$
%exist. 
Given the transition rates ${\bf W}$, the  
stationary probabilities ${\bf p}^{{\rm st}}$ (the diagonal elements of the
reduced density matrix of the dot array) can be found from a 
master equation. By making use of the conservation of probability,
we rewrite the master equation in the form $\tilde {\bf W} {\bf p}^{{\rm st}}=
{\bf v}$, where $\bf \tilde W$ is identical to $\bf W$ but with 
one (arbitrarily chosen) row $\chi_0$ being replaced with 
$(\Gamma,...,\Gamma)$. The vector $\bf v$ is defined as
$v_\chi=\Gamma \delta_{\chi \chi_0}$.
The current and shot noise are given by:
\begin{equation}
  I = {e\over 2\hbar} {\bf e}^T {\bf W}^{I}
  {\bf p}^{{\rm st}}
\label{eq:I1}
\end{equation}
\begin{equation}
  S = {e^2\over \hbar} {\bf e}^T \left( 
    {\bf W}^{II} {\bf p}^{{\rm st}} + {\bf W}^{I} 
    {\bf P} {\bf W}^{I} {\bf p}^{{\rm st}} \right).
\label{eq:S1}
\end{equation}
%The bold face indicates matrix notation related to the eigenstate
%labels $\chi$ (for the $N$ dots system there are $4^N$
%different states). 
The vector $\bf e$ is defined as $e_\chi = 1$ for 
all $\chi$ and the  matrices ${\bf W}^{I} ({\bf W}^{II})$ denote the 
transition rates with one (two)
current vertex replacing one (two) tunneling vertex due to $H_{\rm T,L}$ or 
$H_{\rm T,R}$. 
The "propagator" $ {\bf P} $ is obtained from 
$ \tilde {\bf W} {\bf P}=  {\bf Q}  \label{def_P} $
with $Q_{\chi' \chi} = (p_{\chi'}^{\rm st}-\delta_{\chi' \chi}) 
(1-\delta_{\chi'\chi_0})$.
%${\bf P}$ is of order $\Gamma^{-1}$, 
%thus leading to a non-vanishing contribution of the second
%part in Eq.~(\ref{eq:S1}) even in lowest (first) order perturbation
%theory in the coupling to the electrodes. 
Further details  of the technique can be found in 
Ref. \onlinecite{thielmann}.
% and need not be repeated here. 

Note that the noise expression Eq. \ref{eq:S1}
consists of two terms:  The first term is associated with  ${\bf W}^{II}$, i.e.
noise diagrams with two current vertices in a single irreducible block. 
This term typically gives a noise contribution of $e I$. 
The second term is due to reducible noise diagrams. 
It involves the "propagator"  ${\bf P}$ of the dot system 
and therefore accounts for the electronic structure and the 
correlations on the central system.
It is the second term that is responsible for the interesting correlation 
physics dicussed below.

{\bf Results}-- 
In the following we discuss current and shot noise for a model of type
Eqs. \ref{hamilton}
with $N=3$ dots and a half filled ground state (i.e. one electron per dot).
We use equal couplings, $\Gamma^L_1 = \Gamma^R_3=2.5{\rm \mu eV}$, so 
%and express all energies in units, such that the total line width  
$\Gamma= \Gamma^L_1 + \Gamma^R_3 = 5{\rm \mu eV}$. 
Our perturbation expansion is valid for temperatures much larger 
than the tunnel couplings.
Throughout this paper, we choose $k_{\rm B} T = 5 \Gamma= 0.025 {\rm meV}$
which corresponds to $T \approx 0.25K$.
The dot system  is characterized by the (uniform) level energy $\epsilon$, 
the intra-dot 'Hubbard' repulsion $U$ and the nearest neighbor charge
repulsion $U_{nn}$, which we present in units of ${\rm meV}$. 
%The parameter sets we use throughout this paper are chosen in such way, 
%that the dot system's ground state is half filled, i.e. a three electron
%state. 
Transport is achieved by applying a bias 
voltage $V_{\rm b}$, which is dropped symmetrically and entirely
at the  electrode-dot tunnel junctions, 
meaning that the energies of the dot states are independent of 
the applied voltage.
% even if the couplings $\Gamma^r_l$ are not symmetric. 
The effects  of polarization due to asymmetric or incomplete voltage drops
%and a possible gate voltage 
are straightforward to anticipate, but would only add unnecessary 
complexity to the results presented below.\\
%An illustration of the transport situation for three dots is shown in the
%inset of Fig. \ref{fig:2D-empty}.
%The Fano factor, which is given by the noise to current ratio, $F=S/2 e I$,
%provides additional information about transport properties, not contained in 
%the current-voltage characteristics alone. Therefore, we are interested in 
%studying its dependence on different couplings to the electrodes, 
%the strength of relaxation, the Coulomb charging energy, etc. 
%in order to make predictions of the importance of 
%those parameters for a given experiment.
We point out that 
%out the importance of 
the three dot system is the simplest system that is not pure interface 
and therefore is the minimal model for a truly non-local 
``artificial molecule''. States which are mainly ``localized'' at the 
interfacial dots are competing in transport 
with states that have their main weight at the middle dot (even though
the Hamiltonian remains fully left $\leftrightarrow$ right symmetric).  This
competition can have dramatic effects in the noise characteristics
that have no equivalent for smaller systems.

\begin{figure}[h]
%\vspace*{-0.3cm}
\centerline{\includegraphics[width=8.5cm]{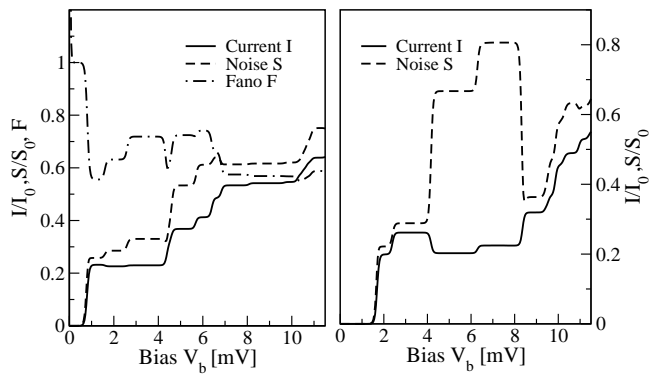}}
\caption{{\it Left panel}: current $I$, shot noise $S$ and Fano factor $F$
  vs. bias voltage for a three-dot chain with 
$k_{\rm B} T=0.025$, $\epsilon=-10 $,
  $t=2 $, $U=12 $ and $U_{nn}=0.2 $   (all energies in units of ${\rm
  meV}$). The current increases stepwise while the noise is sub-Poissonian for
  the all bias larger than the sequential tunneling threshold.
{\it Right panel}: current $I$ and noise $S$ for same parameters as on the
  left, except that  $\epsilon=-10.8 $, $U_{nn}=5 $. 
Above $V_b \sim 4{\rm mV}$ a region of strongly enhanced noise appears, 
whereas the current behaves in a similar manner as in the left panel.
%resulting in a  Fano Factor (not shown here) much larger  than unity. 
All current and noise curves are normalized to 
$I_{\rm  0}=(e/2\hbar)\Gamma\sim 60 {\rm pA}$ and 
  $S_{\rm 0}=(e^2/2\hbar)\Gamma \sim 10^{-29} {\rm A^2/Hz}$, respectively.
}
\label{fig:subpoissonian}
\end{figure}

We first consider a situation where the non-local interaction $U_{nn}$ 
is small ( $U_{nn} = t/10$) and obtain
the typical behavior for a fully symmetric system, see left panel of
Fig. \ref{fig:subpoissonian}. The current rises (mostly) 
monotonically in steps, the
noise also shows steps, but must not increase monotonically
(thermally broadened peaks around the steps are also possible). 
The Fano factor $F=S/2 e I$ will fall
between values of $1$ (Poissonian noise) and $1/2$  for biases larger than all 
excitation energies  (symmetric double barrier noise), 
though in general it will not fall
with a monotonous dependence on bias. At small bias, $eV_b \ll k_{B}T$, the
noise is dominated by thermal noise, leading to a divergence \cite{loss} 
of the Fano factor.

\begin{figure}[h]
%\vspace*{-0.8cm}
\centerline{\includegraphics[width=8.5cm]{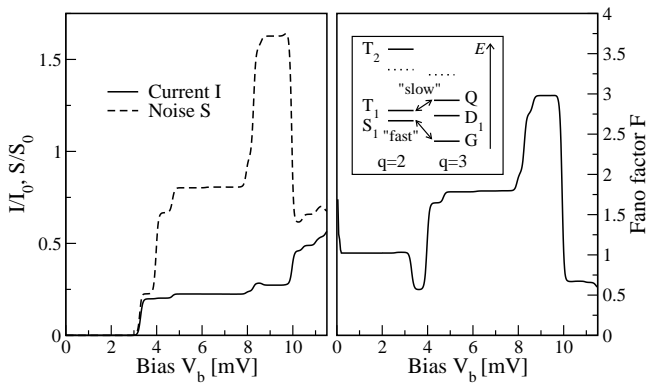}}
\caption{Current $I$ and shot noise $S$ vs. voltage for three dots 
$k_{\rm B} T=0.025$, $t=2$,
  $\epsilon=-10$,  $U=12$ , $U_{nn}=5$. The noise is strongly enhanced in 
absolute magnitude above $V_b\sim 4{\rm mV}$, while the current only slightly
  increases, leading to a Fano factor $F>1$. 
This is due to a competition of "fast" and  "slow" transport channels.
The noise scales like $(U_{nn}/t)^2$ in this regime, while the current 
saturates as $t$ is lowered.
Above $V_b\sim 10{\rm mV}$ "normal" behavior resumes, as the "slow"
  channel of transport is "cut short" (see text). 
%  The current and noise curves are normalized to $I_{\rm max}=(e/\hbar)2\Gamma$ and 
%  $S_{\rm max}=(e^2/\hbar)2\Gamma$, respectively.
}
\label{fig:3D-half-filled}
\end{figure}

If now the non-local Coulomb repulsion $U_{nn}$ is increased 
so that $U_{nn}>t$ we 
observe (right panel of Fig. \ref{fig:subpoissonian}) the usual
stepwise increase of current and noise except for a bias region 
beyond the sequential tunneling threshold, where the
noise is strongly enhanced, accompanied with a decrease in the current,
signaling negative differential conductance (NDC). 
%A similar NDC phenomenon is known from multi-level 
%single quantum dots with asymetric couplings\cite{bulka,thielmann}
%however the noise in that case is quite different. 
The noise enhancement (and NDC) %in the present case 
is the result of the occupation of the quadruplet state $Q$ 
(see inset of Fig. \ref{fig:3D-half-filled} and discussion
below) that, due to the nature of its wave function, contributes a 
''slow'' channel of transport competing with other "fast" channels.

It is important to note that the strong enhancement of the
noise does {\it not require} NDC. 
%The origin of the super poissonian noise
%is connected to the nature of the states wave functions and spin. 
As an example, in Fig. \ref{fig:3D-half-filled} we show
current, noise and Fano factor for the same parameters as for
Fig. \ref{fig:subpoissonian}, right panel, 
except for a different $\epsilon=-10$.
%where the non-local interaction $U_{nn}$ dominates over the hopping $t$.
The current shows generic behavior, 
i.e. stepwise increase and only a tiny NDC around $V_b=8.5 {\rm mV}$. 
The noise, however, is tremendously enhanced,  
with the Fano factor $F > 1$ indicating its super-Poissonian nature.
%For the Coulomb blockade region, the explanation is
%equivalent to the one for Fig. \ref{fig:3D-zero-Unn} with the added
%complexity of having now a variety of excited states available,
%depending on the applied bias. 
While $F>1$ in itself is not uncommon in the sequential tunneling regime
\cite{ferro_leads,bulka,thielmann}, 
often $F>1$ is achieved by a suppression of
transport in which the current is more suppressed than the noise.
Here, the current is not suppressed, but the noise itself is enhanced 
 in absolute magnitude over a large bias range, before recovering 
``normal'' behavior  beyond a bias of $V_b=10 {\rm mV}$.

For such a noise behavior, firstly, strong
electron interactions ($U, U_{nn} \gg T$) 
are needed to have the various states compete in transport.
Secondly, the outcome of this competition is determined 
by the wave functions of the competing states that effectively generate 
state dependent tunneling transition rates. 
Finally, the total spin %${\bf \cal S}$ 
of the states in question can differ by more than the electron spin $1/2$ , so
some energetically and spatially possible transition rates vanish due to spin 
selection rules. In the present case, the dominance 
of the non-local interaction $U_{nn}$ over the hopping $t$,
leads to a strong modification in the spatial
distribution of the relevant  many-body wave functions as compared to the
$U_{nn}\ll t$ case. 
Let us consider the 9 states with total spin $1$ %${\bf \cal S}=1$ 
in the $q=2$ charge sector that split into three triplets (only $T_1$,$ T_2$ shown
in the insert of Fig. \ref{fig:3D-half-filled}. 
If  $U_{nn} < t$, the lowest triplet ($T_1$) will prefer to have
electrons on the middle dot, to maximize the kinetic energy. On the other
hand, if  $U_{nn} > t$ it prefers to have one electron
each on the leftmost and the rightmost dot, thus minimizing 
both intra-dot and inter-dot Coulomb repulsion. This change in the nature of 
the lowest triplet  wave function is crucial for the "noisy" transport. 
An equally fundamental role plays the total spin $3/2$ quadruplet, the 
second excited state in the  $q=3$ charge sector. Due to spin selection rules
these quadruplet states can only have tunneling transitions 
to the triplet states of the $q=2$ (or $q=4$) charge sectors. 
The transitions between the lowest triplet 
and the quadruplet form a "slow" channel of transport. The transition rate 
between them is suppressed by a factor $\sim (t/U_{nn})^2$, 
as the tunneling can happen only at the interface dots, but the wave 
functions of the triplet and 
quadruplet mainly differ by an electron in the middle dot. The current
effectively  alternates between "fast" tunneling sequences  between the well
connected doublet ($D_1$, $G$) and singlet states ($S_1$), and the "slow" sequences between the 
lowest triplet ($T_1$) and quadruplet state ($Q$), leading to
the super-Poissonian noise in the corresponding bias regime.

In Fig. \ref{fig:3D-half-filled}, 
on the first plateau for $3.3 {\rm mV} < V_b < 4 {\rm mV}$ 
with sub-Poissonian noise, transport is mainly achieved by tunneling events 
in which the dot system alternates between the ground state $G$  
with charge $q=3$, and the lowest singlet $S_1$ with charge $q=2$. 
% and triplet states $T_1$ 
The  states  $S_1$ and the lowest triplet $T_1$ ($q=2$) 
are only split by the small "exchange energy" of size 
$J\sim t^2/U \sim 0.33{\rm meV} $ 
and the first excited doublet $D_1$  ($q=2$) falls
energetically in between  $S_1$ and  $T_1$, see the inset in the right panel
of Fig. \ref{fig:3D-half-filled}. Therefore, 
as soon as the sequential tunneling threshold 
$V_{\rm seq} = 2(E_{S_1}-E_G)/e \sim 3.3 {\rm mV} $ is overcome, 
the states $T_1$ and $D_1$ also participate in transport, 
though their occupation turns out to be small (less than 1 percent). 
Since  $V_{\rm seq} > 2(E_{Q}-E_{T_1})/e $ even the quadruplet
$Q$ is occupied, though with the same low probability as the triplet $T_1$ 
\cite{note_fig1}. 
However, above $ V_b = 2(E_{T_1}-E_G)/e = 4 {\rm mV}$
the transport becomes noisy,  with a Fano factor $F > 1$, 
though the current itself still increases by about 2 percent (not visible in
the figure). 
When direct transitions between the ground state $G$  and lowest triplet $T_1$
become possible,  the occupation in $T_1$, $D_1$ and $Q$ increases at the cost
of $G$ and $S_1$, until each state has equal probability. Due to their larger 
spin multiplicity this means that the triplet and the quadruplet states 
compose about $7/12$ of the total probability.
%\begin{figure}[h]
%\vspace*{-0.8cm}
%\centerline{\includegraphics[width=7.5cm]{graphics/E1.15.eps}}
%\caption{Current $I$ and shot noise $S$ vs. voltage for three dots 
%$k_{\rm B} T=0.025$, $t=2$,
%  $\epsilon=-11.5 $,  $U=12$, $U_{nn}=5$.
%The enhanced noise can appear even at the first sequential plateau.
%  The current and noise curves are normalized to $I_{\rm max}=(e/\hbar)2\Gamma$ and 
%  $S_{\rm max}=(e^2/\hbar)2\Gamma$, respectively.
%}
%\label{fig:3D-anion}
%\end{figure}
For even larger bias, states in the $q=4$ charge sector come into play, which
leads to several smaller features, until at about $V_b=8 {\rm mV}$  
the quadruplet becomes even more occupied via the lowest triplet in the 
$q=4$ charge sector and the noise and the Fano factor reach their maximum. 
The maximum value of noise (and Fano factor) behave as $(U_{nn}/t)^2$, whereas
the current value at the corresponding plateau saturates as the hopping
$t$ is decreased. This is due to the homogeneity of our Hamiltonian: site 
disorder, i.e. dot-dependent $\epsilon$, will lead to a 
decreasing current  with decreasing $t$ \cite{elattari}. 
The size of $t$, however, is restricted to the
condition $t \gg \Gamma^r$, which is needed for the division of the total
system into a coherent dot system with perturbative coupling to electrodes to
make sense. 

At $V_b= 10 {\rm mV}$
transitions between the quadruplet and the
second triplet ($T_2$) with $q=2$ become possible. 
In contrast to the lowest triplet $T_1$, 
the states of the second triplet  (as well as the quadruplet states) do not
benefit from the hopping $t$, therefore the structure of their wave functions 
is independent of both the interactions $U, U_{nn}$ and the hopping
$t$, only the actual energy varies ($E_{T_2} = 2\epsilon +U_{nn}$, $E_Q =
3\epsilon +2 U_{nn} $). The second triplet is spread uniformly
over the three dots, its transition rate with the quadruplet is not suppressed 
with  $t/U_{nn}$. Therefore, the slow channel between the $T_1$ triplet 
and the quadruplet is "cut short", the transport becomes more homogeneous
and the Fano factor drops below unity.
% indicating sub-Poissonian noise.

We conclude with a few remarks: 1) As the above effect relies on 
spin quantum numbers we believe  that it will be robust to standard relaxation
processes involving phonon and photon emission. A strong magnetic 
field will modify the details, but not the generic behavior of the transport. 
%2) While we considered 
%parameters such that $q=3$ and $q=2$ charge sectors partiticipate in
%transport, by application of a gate voltage the same effect could be 
%achieved with the  $q=3$ and  $q=4$ charge sectors. 
2) With a change of 
$\epsilon$ (gate voltage), i.e. $\epsilon=-11.5$,
the enhancement of shot noise can be achieved even at the first plateau. 
Although the details of how each state exactly contributes to the transport 
at a given bias are changed, the main mechanism 
(the slow triplet-quadruplet channel) to the enhanced noise is the same.
3) For $U_{nn} \ll t$ a similar spatial distribution of 
the lowest triplet wave function  can be achieved by raising the 
level position $\epsilon_2$ of the middle dot more than $t$ above the 
$\epsilon$ of the interfacial dots. However, such a level detuning will 
strongly reduce both current and shot noise, similar to the "local" models of 
Refs. \onlinecite{bulka,thielmann}. 
%While the  Fano factor will 
%indeed be somewhat larger unity, the usefulness of achieving 
%super-Poissonian noise in this way is rather limited.
4) In the Coulomb blockade region and
close to the sequential tunneling threshold we often find
super-Poissonian noise. This enhancement is due to
the thermal occupation and following sequential depletion of 
excited states that lead to small cascades of tunneling events interrupted 
by long (Coulomb) blockages, resulting in a noisy current. Since the origin
of this effect is completely different to the one discussed above 
we refer the reader to the work of Belzig and co-workers
\cite{belzig1,belzig2} who have recently discussed this effect for systems
restricted to a singly occupied ground state in some detail. 
The effect can be even more pronounced in more complex systems and 
ground states with higher charge $q$ \cite{jasmin-2qd} as the number of 
available excited states typically increases with $q$.

In summary, we have discussed the interplay of Coulomb 
interactions and non-locality in fully symmetric models 
for a chain of quantum dots.
We find super-Poissonian shot noise %in a bias region 
above the sequential
tunneling threshold where the electronic current is  not suppressed. 
The Fano factor becomes large in this regime, because the shot noise 
itself is strongly enhanced.
% becomes much larger than in transport through single quantum dots. 
This effect is caused by a competition of "slow" and "fast" transport channels
due to the differing spatial structure and total spin 
of the  participating many-body states.
%the competition of several many-body states with 
%differing spatial structure and total spin.
We expect the effect to be robust, thus offering a way to study shot noise 
experimentally in chains of lateral quantum dots in semiconducting 
heterostructures.

{\em Acknowledgments.}
We enjoyed interesting and helpful discussions with J. K\"onig
as well as financial support by the DFG via the Center for 
Functional Nanostructures.
% and the Emmy-Noether program.


\begin{thebibliography}{99}
\bibitem{note_fig1} For the parameters of Fig. \ref{fig:subpoissonian}, right
 panel, transport in the first plateau is exclusively between $G$ and $S_1$. 
On the second  plateau $T_1$ and $D_1$ join in and obtain equal occupation. 
Only on the third plateau the bias suffices for the quadruplet $Q$ to be 
occupied, leading to NDC and enhanced noise.
\end{thebibliography}
\end{document}